\documentclass[aps,superscriptaddress,showpacs,nofootinbib,floatfix,twocolumn]{revtex4}
\usepackage{epsfig,bm,feynmf}
\usepackage{graphics}
\usepackage{amsmath}
\usepackage{mathrsfs}
\usepackage[normalem]{ulem}  
\usepackage[dvips]{color} 

\renewcommand\sout{\bgroup \color{red} \ULdepth=-.5ex \ULset}

\begin{document}
\author{Yunpeng Liu}
\email{yunpeng.liu@tju.edu.cn}
\affiliation{Department of Physics, Tianjin University, Tianjin 300072, P.R. China}

\author{Che Ming Ko}
\email{ko@comp.tamu.edu}
\affiliation{Cyclotron Institute and Department of Physics and Astronomy, Texas A$\&$M University, College Station, Texas 77843, USA}%

\title{Thermal production of charm quarks in heavy ion collisions at  Future Circular Collider}
\pacs{25.75.-q, 25.75.Cj, 12.38.Mh}

\begin{abstract}
By solving the rate equation in an expanding quark-gluon plasma, we study thermal production of charm quarks in central Pb+Pb collisions at the Future Circular Collider. With the charm quark production cross section taken from the perturbative QCD at the next-to-leading  order, we find that charm quark production from the quark-gluon plasma can be appreciable compared to that due to initial hard scattering between colliding nucleons.  
\end{abstract}

\maketitle
It is now widely accepted that a quark-gluon plasma (QGP) is produced in  heavy ion collisions at the Relativistic Heavy Ion Collider (RHIC) and the Large Hadron Collider (LHC).  Among the many signals for the QGP  is the suppressed production of quarkonia as a result of color screening and dissociation in the produced QGP~\cite{Matsui:1986dk}. Although regeneration of quarkonia from QGP is possible~\cite{Thews:2000rj, Grandchamp:2001pf, Yan:2006ve}, its contribution is not particularly important as the number of heavy quarks is small, especially at the RHIC energy, 
 and they are mainly produced in the primordial  
scattering of colliding nucleons instead of from the QGP due to their large masses. 
However, with increasing  collision energy,  thermal production of charm quarks becomes  possible at LHC, although the effect is not  large~\cite{Levai:1994dx, Levai:1997bi,Zhang:2007dm,Uphoff:2010sh}.  With the Future Circular Collider (FCC)  that is being discussed for Pb+Pb collisions at $\sqrt{s_{\rm NN}}=39$\ TeV, it becomes of interest to know if the  order of magnitude higher collision energy than currently available  at LHC would lead to a substantial thermal production of charm quarks in these collisions. In this short report, we  calculate the thermal charm yield in heavy ion collisions at FCC energy based on a boost invariant expanding QGP and a kinetic equation for charm quark production using the charm production cross section from the pQCD at the next-to-leading  order~\cite{Nason:1987xz, Zhang:2007dm}.
\par
The kinetic equation for charm quark production in a QGP can be written as~\cite{Zhang:2007dm}
\begin{eqnarray}
   \partial_{\mu}(\rho_c u^{\mu}) &=& R\left[1-\left(\frac{\rho_c}{\rho_c^{\rm eq}}\right)^2\right],
   \label{eq_1}
\end{eqnarray}
where $\rho_c$, $u^{\mu}$\ and $R$\ are, respectively, the local number density of charm quarks, the local 4-velocity, and the thermal production rate of charm quarks, and
\begin{eqnarray}
   \rho_c^{\rm eq}&=& 
   N_{\rm deg}\int \frac{d{\bf p}}{(2\pi)^3}\frac{1}{e^{E/T}+1}\\
   &\approx&
   N_{\rm deg}\int \frac{d{\bf p}}{(2\pi)^3}e^{-\sqrt{ {\bf p}^2+m_c^2}/T}
\end{eqnarray} 
is the  equilibrium density of charm quarks, with the charm quark mass $m_c=1.3$\ GeV and the number of degree of freedom $N_{\rm deg}=6$. Because the charm quark production rate $R$\ is negligible at low temperature~\cite{Nason:1987xz, Zhang:2007dm} as  observed in heavy ion collisions at RHIC~\cite{Adamczyk:2014uip},  thermal production is important only at the early stage of the  produced QGP when the longitudinal expansion is dominant. Therefore, we assume boost invariance and neglect the transverse expansion of the  QGP by  taking $u^{\mu}=(\cosh y, 0, 0, \sinh y)$\ with $y=\textrm{arctanh}(z/t)$\ being the rapidity.  Eq.~(\ref{eq_1}) can then be simplified to 
\begin{eqnarray}
   \partial_{\tau}\sigma_c &=& \tau R\left[1-\left(\frac{\sigma_c}{\sigma_c^{\rm eq}}\right)^2\right],
   \label{eq_sigma_c}
\end{eqnarray}
where we have introduced the area density $\sigma_c\equiv dN_c/(dyd{\bf x}_T)=\tau \rho_c$ of charm quarks, which would be constant if there were no thermal production or annihilation. Here $\tau=\sqrt{t^2-z^2}$\ and ${\bf x}_T$ are the proper time and tranverse coordinates, respectively. In the above, both the thermal  production rate $R$\ and the equilibrium area density $\sigma_c^{\rm eq}$\ depends on the local temperature $T$ of the QGP. 
\par
For the thermal charm quark production rate, we take it from Ref.~\cite{Zhang:2007dm} based on the charm quark production cross sections from the next-to-leading-order QCD calculation given in Refs.~\cite{Nason:1987xz,Nason:1989zy,Beenakker:1988bq,Beenakker:1990maa} and using thermal masses $m_q=gT/\sqrt{6}$ and $m_g=gT/\sqrt{2}$ for quarks and gluons, respectively. Specifically, it includes charm quark production from the leading-order processes $q+\bar q\to +c+\bar c$ and $g+g\to c+\bar c$ as well as the next-order processes $q+\bar q\to c+\bar c+g$ and $g+g\to c+\bar c+g$ and the interferences between the leading-order processes with their virtual corrections due to vertex corrections and self energy insertions. The processes $g+q\to c+\bar c+q$ and $g+\bar q\to c+\bar c+\bar q$ are, however, neglected due to their smaller cross sections. To facilitate our calculations, we parameterize the thermal charm quark production rate shown in Fig.3 of Ref.~\cite{Zhang:2007dm} as
\begin{eqnarray}
   \log_{10}R=\sum_{n=0}^5 a_nT^n
\end{eqnarray}
with $a_0=-18.2327$, $a_1=110.9367$, $a_2=-319.9090$, $a_3=506.6754$, $a_4=-413.4846$, and $a_5=136.0222$, where $T$\ is in GeV and $R$\ is in $c/$fm$^4$.
\par
Since only the longitudinal expansion is considered, the time evolution of the  QGP can be approximately described by entropy conservation
\begin{eqnarray}\label{entropy}
   s({\bf x}_T, \tau)&=& \frac{\tau_r}{\tau}s({\bf x}_T,\tau_r),
\end{eqnarray}
where $s$\ is the local entropy density  and is assumed to be known at some given time $\tau_r$.   Because of the large collision energy at FCC, we further assume that the entropy density  is proportional to the number of binary collisions $n_{\rm coll}({\bf x}_T)=\sigma_{pp}^{\rm in}T_A({\bf x}_T)T_B({\bf x}_T)$ between the two colliding nuclei, where $\sigma_{pp}^{\rm in}$\ is the proton-proton inelastic cross section and $T_A$ ($T_B$) is the thickness function of nucleus A (B). Therefore, Eq.(\ref{entropy}) can be rewritten as
\begin{eqnarray}
   s({\bf x}_T, \tau)&=& \frac{\tau_r T_A({\bf x}_T)T_{B}({\bf x}_T)}{\tau T_A({\bf 0})T_B({\bf 0})} s({\bf 0}, \tau_r)
\end{eqnarray}
in terms of the entropy density in the center of the QGP at time $\tau_r$.  

Through the equation of state of produced hot dense matter, $s({\bf x}_T,\tau_r)$ is related to the energy density $\epsilon({\bf x}_T,\tau_r)$ and can be determined from the transverse energy $dE_T/dy$ via
\begin{eqnarray}
\int d{\bf x}_T\epsilon(s({\bf x}_T,\tau_r))=\frac{1}{\tau_r}\frac{dE_T}{dy}.
\end{eqnarray}

 According to Ref.~\cite{Chatrchyan:2012mb}, the energy dependence of the transverse energy measured in heavy ion collisions  can be parametrized as
\begin{eqnarray}
   \frac{dE_T}{d\eta}&=& A\left(\frac{\sqrt{s_{\rm NN}}}{\sqrt{s_{\rm NN}^0}}\right)^{0.4}\frac{N_{\rm part}}{2},
\end{eqnarray}
with $A=0.46$\ GeV and $\sqrt{s_{\rm NN}^0}=1$\ GeV. The  number of participants, $N_{\rm part}$, in above equation can be obtained from  the $p+p$\ inelastic cross section using the parametrization~\cite{Zsigmond:2012vc} 
\begin{eqnarray}
   \sigma_{pp}^{\rm in}(\sqrt{s})&=& \sigma_0\ln\frac{\sqrt{s}}{\sqrt{s_0}},
\end{eqnarray}
with $\sigma_0=8.2$\ mb and $\sqrt{s_0}= 1.436$\ GeV. With an  inelastic cross section $\sigma_{pp}^{\rm in}=84$\ mb at 39 TeV, we have  $N_{\rm part}=408.5$ for central Pb+Pb collisions at same energy, resulting in a  transverse energy $dE_T/dy=6447$\ GeV  if we take $\tau_r=5$ fm.  
\par
\begin{figure}[!hbt]
    \centering
    \includegraphics[width=0.45\textwidth]{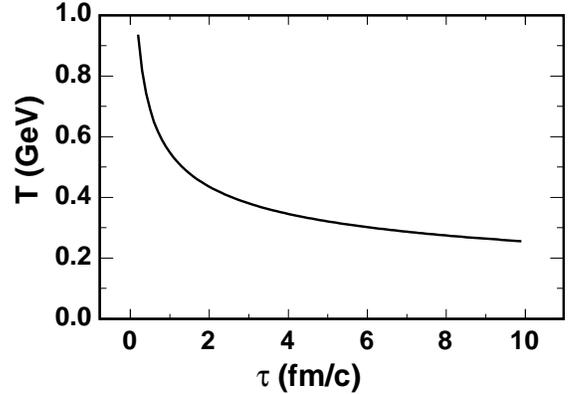}
    \caption{Time evolution of local temperature $T$\ at ${\bf x}_T=0$.}
    \label{fg_T}
 \end{figure}
\begin{figure}[!hbt]
    \centering
    \includegraphics[width=0.45\textwidth]{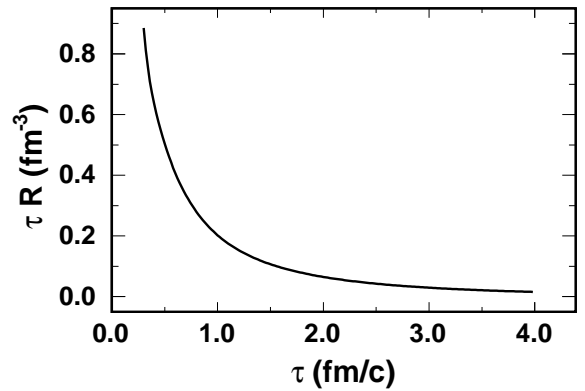}
    \caption{Time evolution of $\tau R$\ at ${\bf x}_T=0$.}
    \label{fg_tauR_xt0}
\end{figure}

Taking the equation of state as an ideal gas of quarks and gluons with masses $m_u=m_d=m_g=0$\ and $m_s=150$\ MeV  for the QGP  and  a resonance gas of hadrons with masses below 2 GeV as well as including a bag constant, which leads  to a first order phase transition at $T_c=165$\ MeV, we have determined the energy density and temperature of the produced medium. In Fig.~\ref{fg_T}, we show the  time evolution of the local temperature at ${\bf x}_T=0$. It is  seen that the temperature is initially about 935 MeV but drops fast at the beginning due to the strong longitudinal expansion and becomes less than $400$\ MeV after $\tau=2.6$\ fm$/c$. As shown in Fig.~\ref{fg_tauR_xt0}, the  production rate $\tau R$  also decreases fast with time and is only important  during the early stage of the  the expanding QGP.  The ratio $\rho_c/\rho_c^{\rm eq}$\ at ${\bf x}_T=0$ is found to increase with time but never exceed 0.42 at $\tau <2.6$\ fm,  indicating that  charm quark annihilation is far  less important than  charm production in the QGP.
\par
 For the initial  charm quark density $\sigma_c(\tau_0)$, it is  estimated by the Glauber model, i.e.,
\begin{eqnarray}
   \sigma_c({\bf x}_T, \tau_0) &=& T_A({\bf x}_T)T_B({\bf x}_T)d\sigma_{pp}^{c\bar{c}}/dy,
\end{eqnarray}
where $\sigma_{pp}^{c\bar{c}}$\ is the charm quark production cross section in $p+p$\ collisions.  From the charm quark production cross section measured in $p+p$\ collisions at $\sqrt{s_{\rm NN}}=2.76$\ TeV, i.e.,   $d\sigma_{pp}^{c\bar{c}}/dy=0.62$\ mb~\cite{Averbeck:2011ga}, we  extrapolate it to $\sqrt{s_{NN}}=39$\ TeV by running PYTHIA~\cite{Sjostrand:2006za, Sjostrand:2007gs} at both energies and obtain the cross section $d\sigma_{pp}^{c\bar{c}}/dy=1.57$\ mb at $\sqrt{s}=39$\ TeV, which leads to the initial quark number $dN_c/dy=48$\  from primordial collisions.
\begin{figure}[!hbt]
    \centering
    \includegraphics[width=0.45\textwidth]{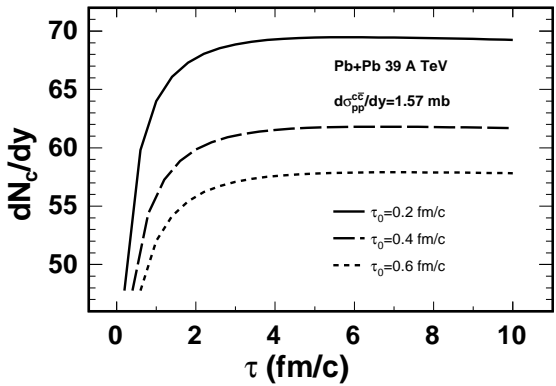}
    \caption{Time evolution of the number of charm quarks with different initial time $\tau_0$.}
    \label{fg_Nc}
\end{figure}
\par
The final yield of charm quarks obtained from solving Eq.~(\ref{eq_sigma_c}) depends on the time $\tau_0$\ when the production of charm quarks starts. It should neither be  much smaller than the formation time of charm quark $1/m_c\sim 0.2$\ fm$/c$\ nor be  much larger than the formation time of QGP.  In Fig.~\ref{fg_Nc}, results from using  $\tau_0=0.2$, $0.4$, and $0.6$\ fm/$c$ are shown. It  is seen that this leads to a  relative enhancement that varies from $21\%$\ to $45\%$  and is thus not negligible. The results of a similar calculation for $\sqrt{s_{\rm NN}}=5.5$\ TeV is from $6\%$\ to $16\%$.
\par
 In summary, we have studied charm quark production from the QGP produced in heavy ion collisions at 39 TeV in future FCC. Using the charm production cross section in quark-anti-quark and quark (anti-quark)-gluon scattering calculated in the next-to-leading order in QCD and assuming that the produced QGP expands boot invariantly, we have found that charm production from the QGP is not negligible. Depending the formation time of the QGP, its contribution can be near to 50\% for a formation time of $\tau_0=0.2$ fm/$c$. Such an enhanced production of charm quarks than that produced from initial hard scattering is expected to have a significant effect on charmonium production in heavy ion collisions at such an energy~\cite{Zhou:2016wbo}.  Work is in progress to study the effect of thermal charm production on the nuclear modification factors for both charm quarks and the charmonia. 
\section*{Acknowledgements}
We thank Andrea Dainese for suggesting this study and helpful discussions. This work was supported by the US Department of Energy under Contract No. DE-SC0015266, the Welch Foundation under Grant No. A-1358, and the NSFC under Grant No. 11547043.
\bibliographystyle{elsarticle-num.bst}
\bibliography{ref}
\end{document}